\documentclass[aps,prb,twocolumn,showpacs,preprintnumbers,amsmath,amssymb,superscriptaddress,10pt]{revtex4-1}%
\usepackage{graphicx}
\usepackage{dcolumn}
\usepackage{amsmath}
\usepackage{color}
\usepackage{bm}

\begin{document}
\title{Nearly itinerant ferromagnetism in CaNi$_2$ and CaNi$_3$}

\author{A. Jesche}
 \email[]{jesche@ameslab.gov}
 \affiliation{The Ames Laboratory, Iowa State University, Ames, USA}
\author{K. W. Dennis}
 \affiliation{The Ames Laboratory, Iowa State University, Ames, USA}
\author{A. Kreyssig}
 \affiliation{The Ames Laboratory, Iowa State University, Ames, USA}
 \affiliation{Department of Physics and Astronomy, Iowa State University, Ames, USA}
\author{P. C. Canfield}
 \affiliation{The Ames Laboratory, Iowa State University, Ames, USA}
 \affiliation{Department of Physics and Astronomy, Iowa State University, Ames, USA}

\begin{abstract}
Single crystals of CaNi$_2$ and CaNi$_3$ were successfully grown out of excess Ca. 
Both compounds manifest a metallic ground state with enhanced, temperature dependent magnetic susceptibility.
The relatively high Stoner factors of $Z = 0.79$ and $Z = 0.87$ found for CaNi$_2$ and CaNi$_3$, respectively, reveal their close vicinity to ferromagnetic instabilities. 
The pronounced field dependence of the magnetic susceptibility of CaNi$_3$ at low temperatures ($T < 25$\,K) suggests strong ferromagnetic fluctuations. A corresponding contribution to the specific heat with a temperature dependence of $T^3$ln$T$ was also observed.
\end{abstract}

\maketitle
\section{Introduction}
Compounds of electropositive elements (e.g. Ca, Ti, La) and transition metals (e.g. Ni, Cu, Cr) have been studied, to a certain extent, to explore their potential for hydrogen storage applications\,\cite{Ishiyama1995, Chen2000}. 
Among these, the members of the Ca-Ni family (CaNi$_2$, CaNi$_3$, Ca$_2$Ni$_7$, and CaNi$_5$) were investigated by means of X-ray diffraction\,\cite{Buschow1974}. 
Cubic CaNi$_2$ and trigonal CaNi$_3$ were found to form CaNi$_2$H$_{3.4}$ and CaNi$_3$H$_{4.6}$ under elevated H-vapor pressures without changing the symmetry of the crystal structure\,\cite{Oesterreicher1980}. 
However, despite these earlier investigations and the simplicity of these binary compounds, no experimental data on physical properties have been reported.
The main reason for this lack of information is the challenging synthesis of these compounds caused by the high reactivity of Ca, which is highly air and moisture sensitive and tends to attack several standard crucible materials and manifests elevated vapor pressures of 1\,bar at $T = 1500^\circ$C (close to the melting point of elemental Ni). 

In this paper we present a method for the growth of CaNi$_2$ and CaNi$_3$ single crystals from Ca-flux.
Measurements of temperature dependent magnetization, electrical resistivity, and specific heat reveal a metallic ground state with an unusual high Stoner-factor, indicating strong ferromagnetic correlations in CaNi$_2$ that are further enhanced in CaNi$_3$.

\section{Experimental}
X-ray powder diffraction (XRD) was performed on ground single crystals using a Rigaku Miniflex diffractometer (wavelength: Cu-$K\alpha_{1,2}$).
Lattice parameters were refined by the LeBail method using GSAS\cite{Larson2000} and EXPGUI\cite{Toby2001}.
Laue-back-reflection patterns were taken with an MWL-110 camera manufactured by Multiwire Laboratories. 
Magnetization measurements were performed using a Quantum Design MPMS.
Electrical resistivity was measured in 4-point geometry using the AC transport option of a Quantum Design PPMS. 
Silver epoxy was used to make electrical contacts on the samples and then cured at $T = 120^\circ$C under air for $\approx$\,20\,min. 
The samples did not visually degrade and XRD measurements revealed no structural changes as a result of this treatment.
Absolute values of the electrical resistivity are approximate due to the poorly defined geometry factor of the samples.
Specific heat was measured by a heat-pulse relaxation method using a Quantum Design PPMS.

\section{Crystal growth}
\begin{figure}
\includegraphics[width=0.45\textwidth]{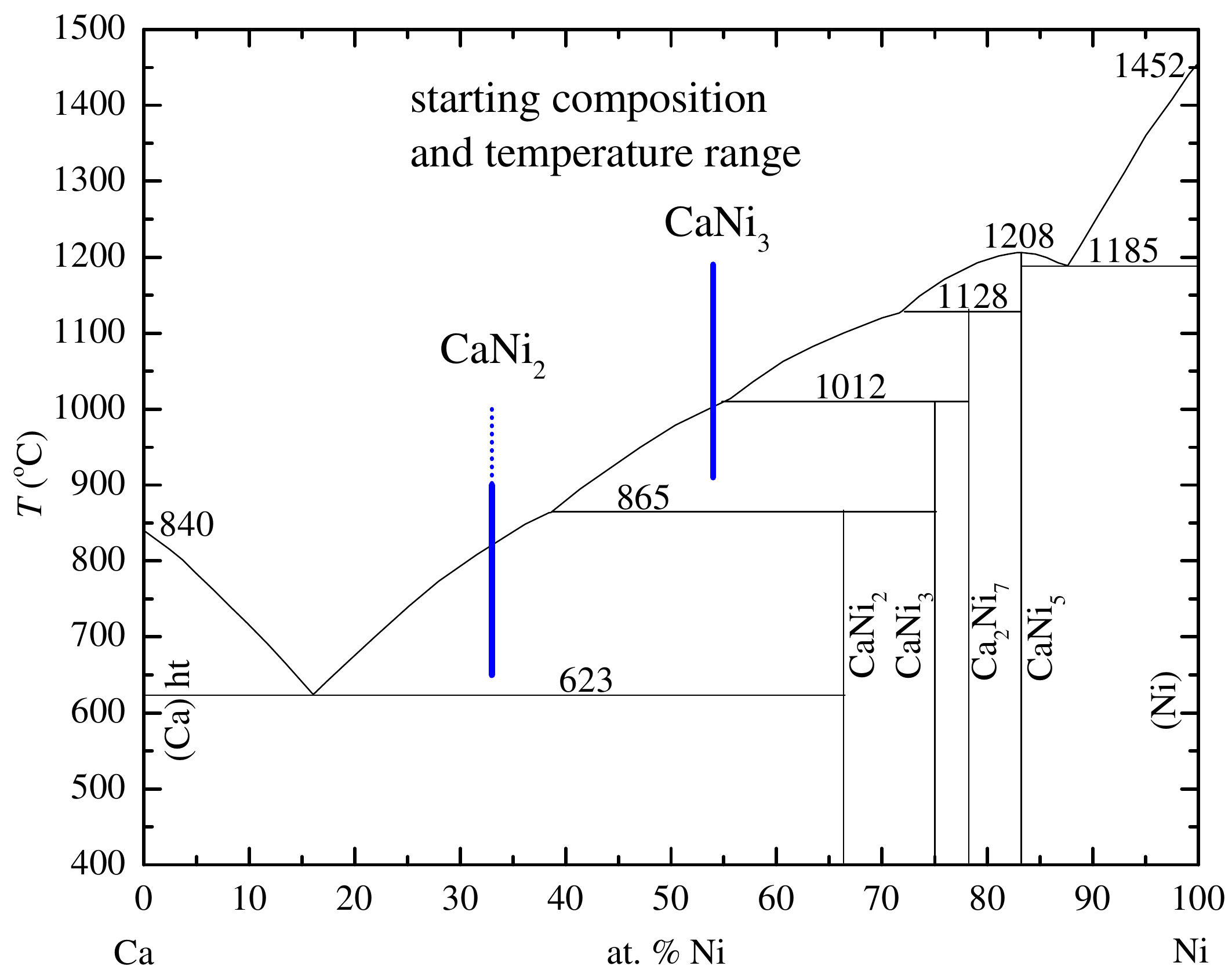}
\caption{(color online) Ca-Ni phase diagram after Notin\,\textit{et\,al.}\,\cite{Notin1990}. Starting compositions and temperature profile for CaNi$_2$ and CaNi$_3$ growths shown with blue lines (the dotted line represents rapid cooling).}
\label{phasediagram}
\end{figure}

Starting materials were Ni-wire (Alfa Aesar, 99.98\% metals basis) and distilled Ca (Ames Laboratory, Metals Developement, 99.98\%).
Best results were obtained by mixing Ca and pieces of Ni-wire in a molar ratio of 67:33 and 46:54 for CaNi$_2$ and CaNi$_3$, respectively, motivated by the published phase diagrams\,\cite{Notin1990,Okomoto1990} (Fig.\,\ref{phasediagram}). 
The mixtures, each with a total mass of roughly 2.5\,g, were packed into a 3-cap Ta-crucible\,\cite{Canfield2001} inside an Ar-filled glove box. 
A combination of laser welding and arc-melting was used to seal the Ta-crucibles under inert atmosphere (0.5\,bar Ar). 

In accordance with the results of Ref.\,\onlinecite{Oesterreicher1980} we found no indications for an attack of Ta-crucibles by Ni.

\begin{figure}
\includegraphics[width=0.45\textwidth]{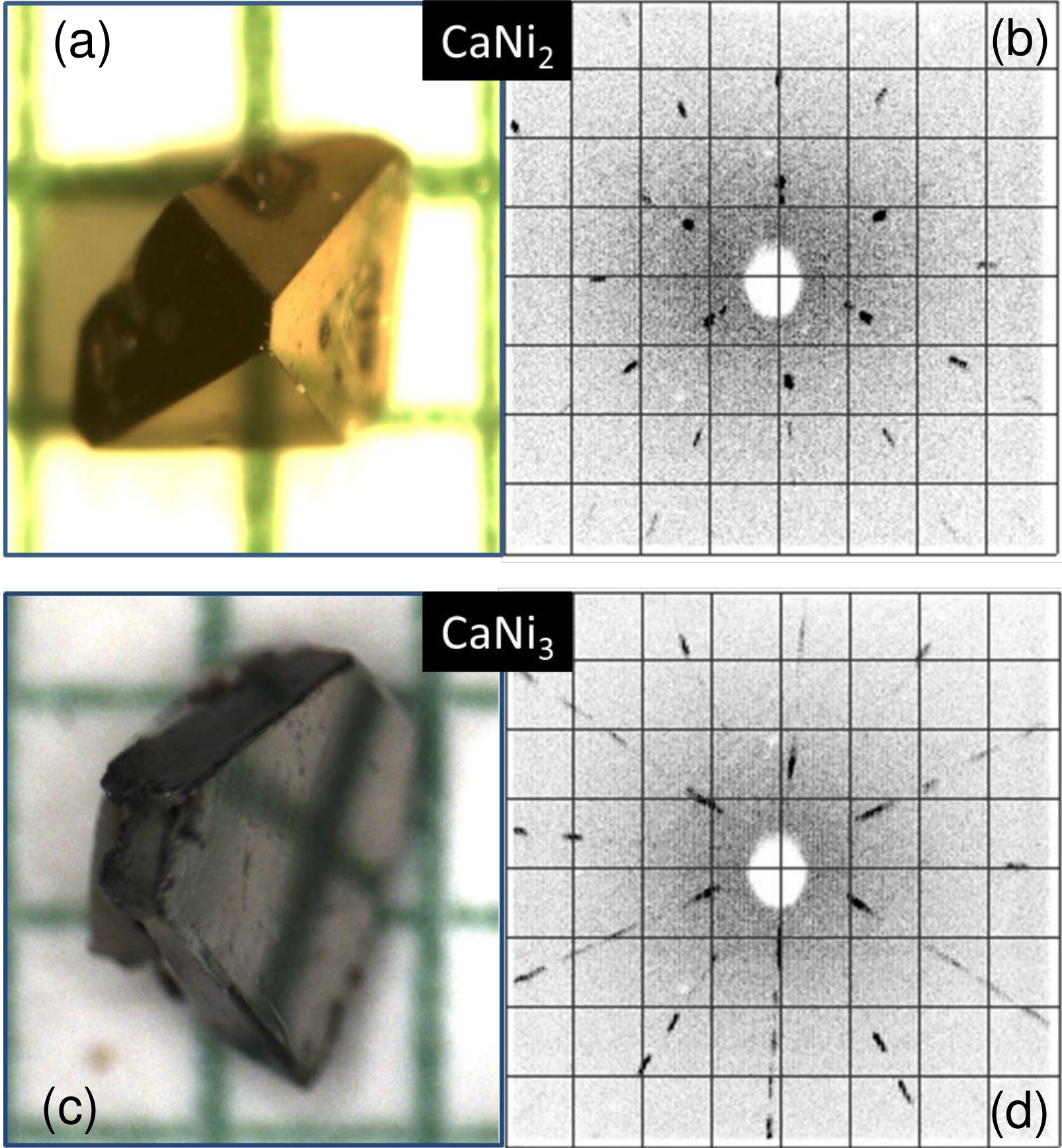}
\caption{(Color online) a) CaNi$_2$ single crystal on a millimeter grid and b) corresponding x-ray Laue back reflection pattern. c) CaNi$_3$ single crystal on a millimeter grid and d) corresponding x-ray Laue back reflection pattern. Both diffraction patterns show the three-fold rotation symmetry of \{111\} and \{001\} directions of their cubic and hexagonal unit cells, respectively.}
\label{bilder+laue}
\end{figure}

\subsection{CaNi$_2$}
The Ca-Ni mixture was heated from room temperature to $T = 1000^\circ$C over 5\,h, cooled to $T = 900^\circ$C within 1\,h, slowly cooled to $T = 650^\circ$C over 50\,h and finally decanted to separate the CaNi$_2$ crystals from the excess liquid. 
Single crystals of octahedral habit with dimensions up to 3\,mm and masses of 25 mg could be obtained (Fig.\,\ref{bilder+laue}a). 
The facets show a three fold rotation symmetry corresponding to \{111\} planes of the cubic lattice as confirmed by Laue back reflection (Fig.\,\ref{bilder+laue}b.)
The triangular hopper morphology seen in the center of some surfaces points to a surface diffusion limited growth that could be further improved best by stirring the melt or rotating the crucible or, to a lesser extent, by decreasing the cooling rate, see e.g. Ref.\,\onlinecite{Elwell1975}.

\subsection{CaNi$_3$}
The Ca-Ni mixture was heated from room temperature to $T = 1190^\circ$C over 6\,h, held at $T = 1190^\circ$C for 1/2\,h, slowly cooled to $T = 910^\circ$C over 32\,h and finally decanted to separate the CaNi$_3$ crystals from the excess liquid.
Plate-like single crystals with lateral dimensions of up to 3 mm and thickness of 0.5 mm could be obtained (Fig.\,\ref{bilder+laue}c).
The threefold crystallographic $\bm c$-axis is oriented perpendicular to the large surface of the plates as confirmed by Laue back-reflection (Fig.\,\ref{bilder+laue}d).

\section{Structural characterization}
After grinding and mounting the sample in an Ar glove box, the powder was covered with capton-foil that was attached to the sample holder using double-faced scotch tape. Once covered in this manner the sample was removed from the Ar glove box. 
The diffraction pattern taken on covered CaNi$_2$ and CaNi$_3$ samples did not change after removing the capton-foil and exposing the powder to air for one week. 
Therefore, both compounds are significantly less air sensitive than elemental Ca. 
However, the development of a field-dependence in the magnetic susceptibility at room-temperature was observed in samples that were stored under air for several weeks indicating the formation of a small ferromagnetic phase presumably due to formation of elemental Ni.

\begin{figure}
\includegraphics[width=0.47\textwidth]{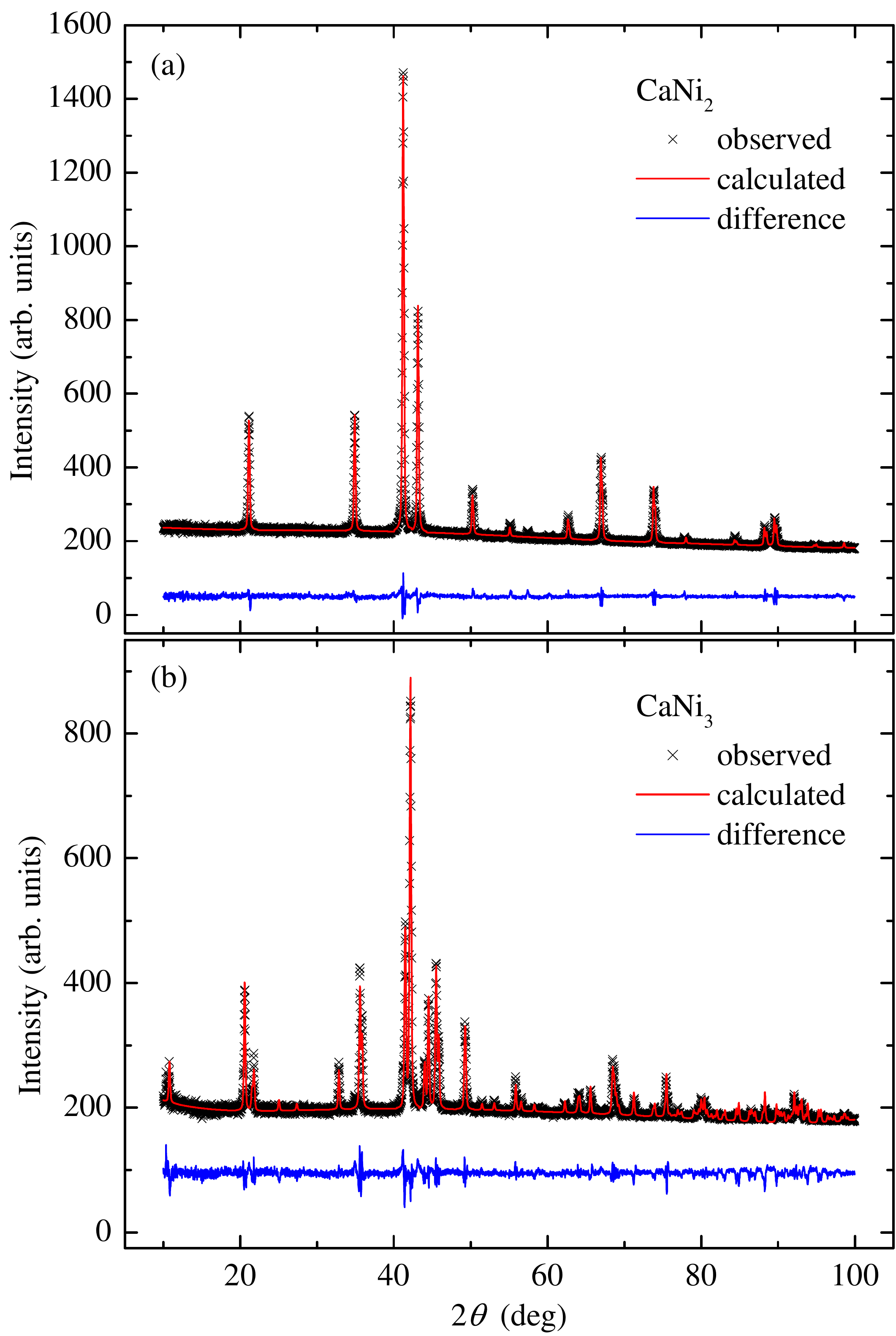}
\caption{(Color online) XRD powder patterns of a) CaNi$_2$ and b) CaNi$_3$ together with fits based on the published cubic and hexagonal structure, respectively.}
\label{diffraktogramme}
\end{figure}

Figure\,\ref{diffraktogramme}a shows the XRD pattern measured on ground CaNi$_2$ single crystals of together with the refined pattern based on the published crystal structure [space group $Fd\bar{3}m$ (227)]. 
The lattice parameter $a = 7.252(6)$\,\AA~is in good agreement with the literature data ($a = 7.251$\,\AA\,\cite{Oesterreicher1980}).

The XRD pattern measured on ground CaNi$_3$ single crystals is plotted in Fig.\,\ref{diffraktogramme}b together with the refined pattern based on the published crystal structure [(space group $R\bar{3}m$ h, (166)].
The lattice parameters of $a = 5.044(3)$\,\AA~and $c = 24.44(9)$\,\AA~are in good agreement with the literature data ($a = 5.052$\,\AA~and $c = 24.45$\,\AA\,\cite{Chen2000}).

\section{Results}

\subsection{Magnetization}

\begin{figure}
\includegraphics[width=0.4\textwidth]{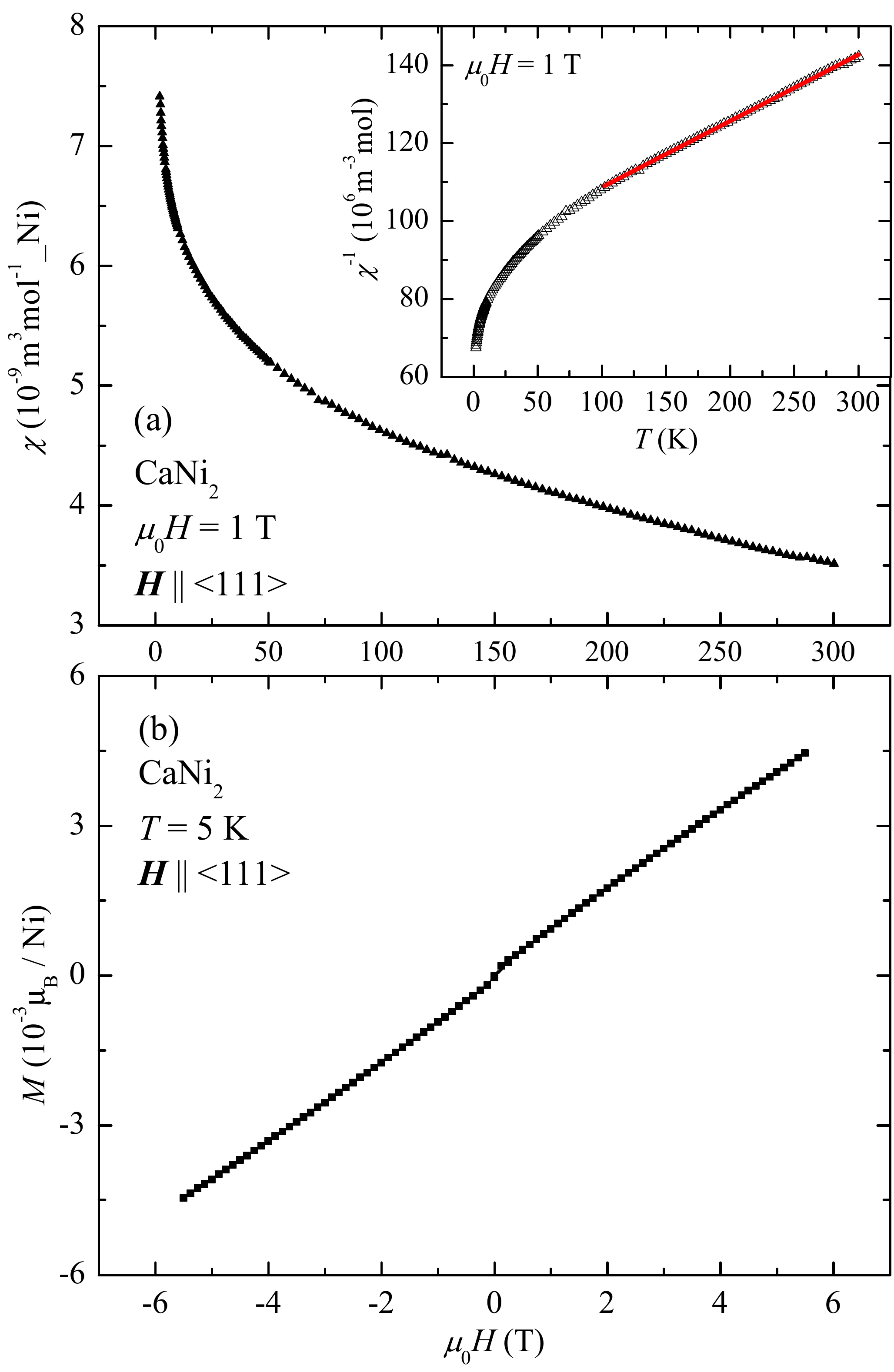}
\caption{(Color online) a) Magnetic susceptibility $\chi = M/H$ per mol Ni of CaNi$_2$. 
The moderate increase of $\chi(T)$ under cooling from 300\,K to $T \approx 20$\,K is followed by a significant increase at lower temperatures. 
The inset shows the inverse susceptibility.
b) Magnetization $M(H)$ as a function of field. 
} 
\label{cani2_mag}
\end{figure}

\begin{figure}
\includegraphics[width=0.47\textwidth]{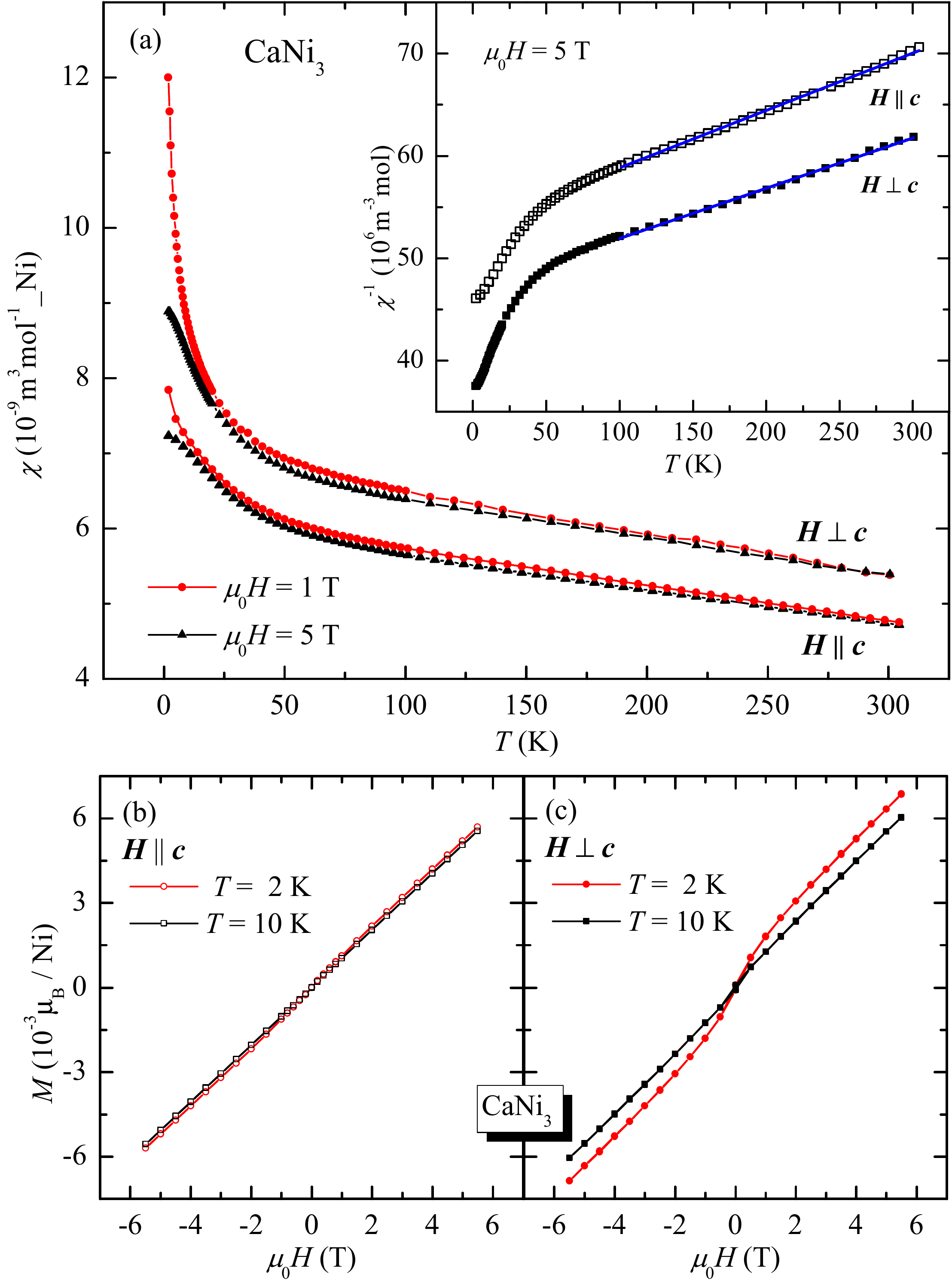}
\caption{(Color online) a) Magnetic susceptibility $\chi = M/H$ of CaNi$_3$ per mol Ni. 
The moderate increase of $\chi(T)$ under cooling from 300\,K to $T \approx 20$\,K is followed by a significant increase at lower temperatures. 
The inset shows the inverse susceptibility.
Magnetization $M(H)$ as a function of field for $\bm H \parallel \bm c$ (b) and $\bm H \perp \bm c$ (c) at $T = 2$ and 10\,K. 
} 
\label{cani3_mag}
\end{figure}

The magnetic susceptibility data $\chi(T) = M/H$ per mol Ni are shown in Fig.\,\ref{cani2_mag}a and Fig.\,\ref{cani3_mag}a for CaNi$_2$ and CaNi$_3$, respectively.
In both compounds $\chi(T)$ is increasing significantly with decreasing $T$ which is in contrast to the $T$-independent behavior expected for a simple metal.
Between $T = 100$ and 300\,K $\chi(T)$ is proportional to $T^{-1}$ manifesting what could be interpreted as a Curie-Weiss-like behavior (insets in Fig.\,\ref{cani2_mag}a and Fig.\,\ref{cani3_mag}a).

For CaNi$_2$, the effective moment that can be inferred from this analysis is $\mu_{\rm eff} = 1.4(1)\,\mu_B$ per Ni and the corresponding Weiss-temperature amounts to a large antiferromagnetic (AFM) value of $\Theta_W = -540$\,K.
Figure\,\ref{cani2_mag}b shows the magnetization as a function of field $M(H)$ in units of $\mu_B$ per Ni.
$M(H)$ at $T = 5$\,K increases in an essentially linear fashion with magnetic field with only a tiny anomaly around $H \approx 0$. 

For CaNi$_3$, the effective moment that can be inferred from this analysis is $\mu_{\rm eff} = 1.95(5)\,\mu_B$ ($\bm H \parallel \bm c$) and $\mu_{\rm eff} = 2.08(5)\,\mu_B$ ($\bm H \perp \bm c$) per Ni.
The Weiss-temperatures amount to large AFM values of $\Theta_W^{\bm H \parallel \bm c} = -950$\,K and $\Theta_W^{\bm H \perp \bm c} = -960$\,K.
As will be discussed below such high Weiss-temperatures are unphysical for intermetallic compounds that remain paramagnetic and indicate that such a local moment treatment of the data is most likely inappropriate.
For $T \geq 25$\,K CaNi$_3$ shows nearly no field-dependence of $\chi(T)$. 
In contrast to the high-$T$ behavior, a pronounced field-dependence of $\chi(T)$ is emerging at $T < 25$\,K - more pronounced for $\bm H \perp \bm c$ - indicating strong ferromagnetic (FM) correlations. 

$M(H)$ of CaNi$_3$ shows an almost perfectly linear field-dependence for $T \geq 10$\,K (for both $\bm H \parallel \bm c$ and $\bm H \perp \bm c$, black squares in Fig.\,\ref{cani3_mag}b,c).
At lower temperatures ($T = 2$\,K) a small deviation from the linear behavior in $M(H)$ is found for $\bm H \parallel \bm c$, whereas a clear curvature forms in $M(H)$ for $\bm H \perp \bm c$ (red circles in Fig.\,\ref{cani3_mag}b,c) in accordance with the field-dependence of $\chi(T)$ at low temperatures.

\subsection{Electrical resistivity}
\begin{figure}
\includegraphics[width=0.45\textwidth]{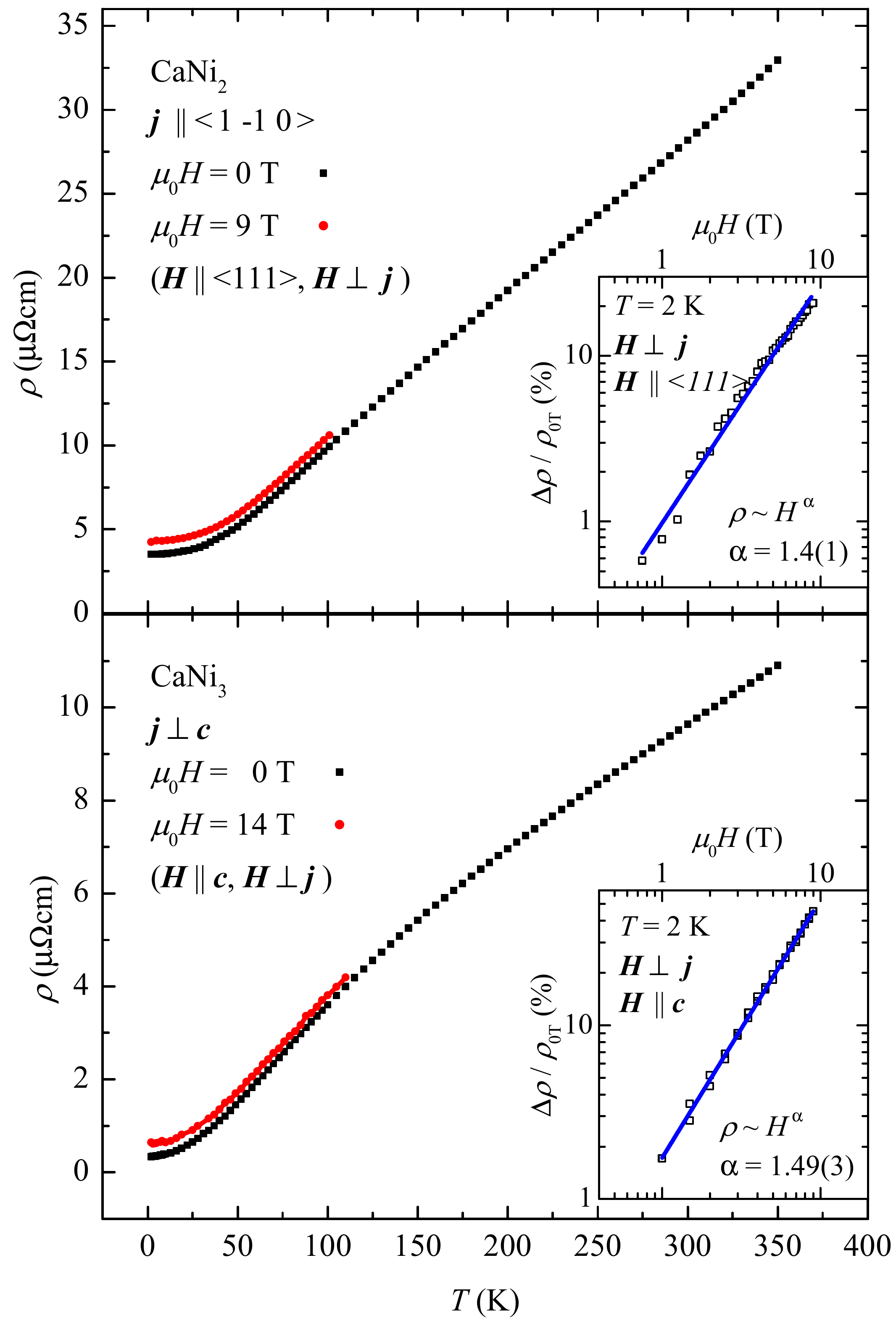}
\caption{
(Color online) Electrical resistivity of a) CaNi$_2$ and b) CaNi$_3$. CaNi$_2$ shows metallic behavior with a linear $T$-dependence over a wide temperature range and a crossover to $T^2$-dependence at low $T$. An additional change of slope is observed for CaNi$_3$ in the region around $T \approx 150$\,K.
(Absolute values are approximate due to the poorly defined geometry factor of the samples.)
Insets: the magnetoresistance $\Delta \rho/\rho_0 = (\rho_H - \rho_0)/\rho_0$ plotted on a double-logarithmic scale at $T = 2$\,K is positive and approaches values of 22\% and 46\% at $\mu_0 H = 9$\,T for CaNi$_2$ and CaNi$_3$, respectively. 
}
\label{rho}
\end{figure}

Figure\,\ref{rho} shows the electrical resistivity of CaNi$_2$ (a) and CaNi$_3$ (b) measured with current flow along $\langle\,1\,\bar{1}\,0\,\rangle$ and along the basal plane, respectively. 
The metallic behavior observed for CaNi$_2$, demonstrated by the linear $T$-dependence for $T > 50$\,K and an approximately quadratic $T$-dependence at low $T$, is modified for CaNi$_3$ by an additional change of slope over a wide $T$-range between $T = 100$ and 200\,K.
Residual resistivity ratios of RRR = $\rho_{300\,K}/\rho_0 = 8$ and 29 for CaNi$_2$ and CaNi$_3$, respectively, are consistent with good quality single crystals.

A positive magnetoresistance $\Delta \rho = \rho_H-\rho_0$ is observed for both compounds with increasing values towards low $T$ (filled, red circles and insets in Fig.\,\ref{rho}, $\bm H \perp \bm j$). 
The field-dependence of $\Delta\rho$ at $T = 2$\,K follows a power-law dependence of $\Delta\rho(H) \sim H^\alpha$ with $\alpha \approx 1.5$ \lbrack 1.4(1) and 1.49(3) for CaNi$_2$ and CaNi$_3$, respectively\rbrack, which differs significantly from the expected value of $\alpha = 2$ for a simple metal.
The values of $\Delta \rho/\rho = 22$\% and 46\% at $\mu_0H = 9$\,T for CaNi$_2$ and CaNi$_3$, respectively, can be regarded as rather high taking into account the comparatively low RRR values when compared to simple metals in the picture of a Kohler plot (see e.g. Ref.\,\onlinecite{Pippard1989}) and are comparable with YAgSb$_2$\,\cite{Myers1999b} but are significantly smaller than the large values of $\Delta\rho/\rho = 5\cdot 10^5$\,\% found in PtSn$_4$\,\cite{Mun2012} (RRR $\sim$ 1000).

\subsection{Specific heat}

\begin{figure}
\includegraphics[width=0.4\textwidth]{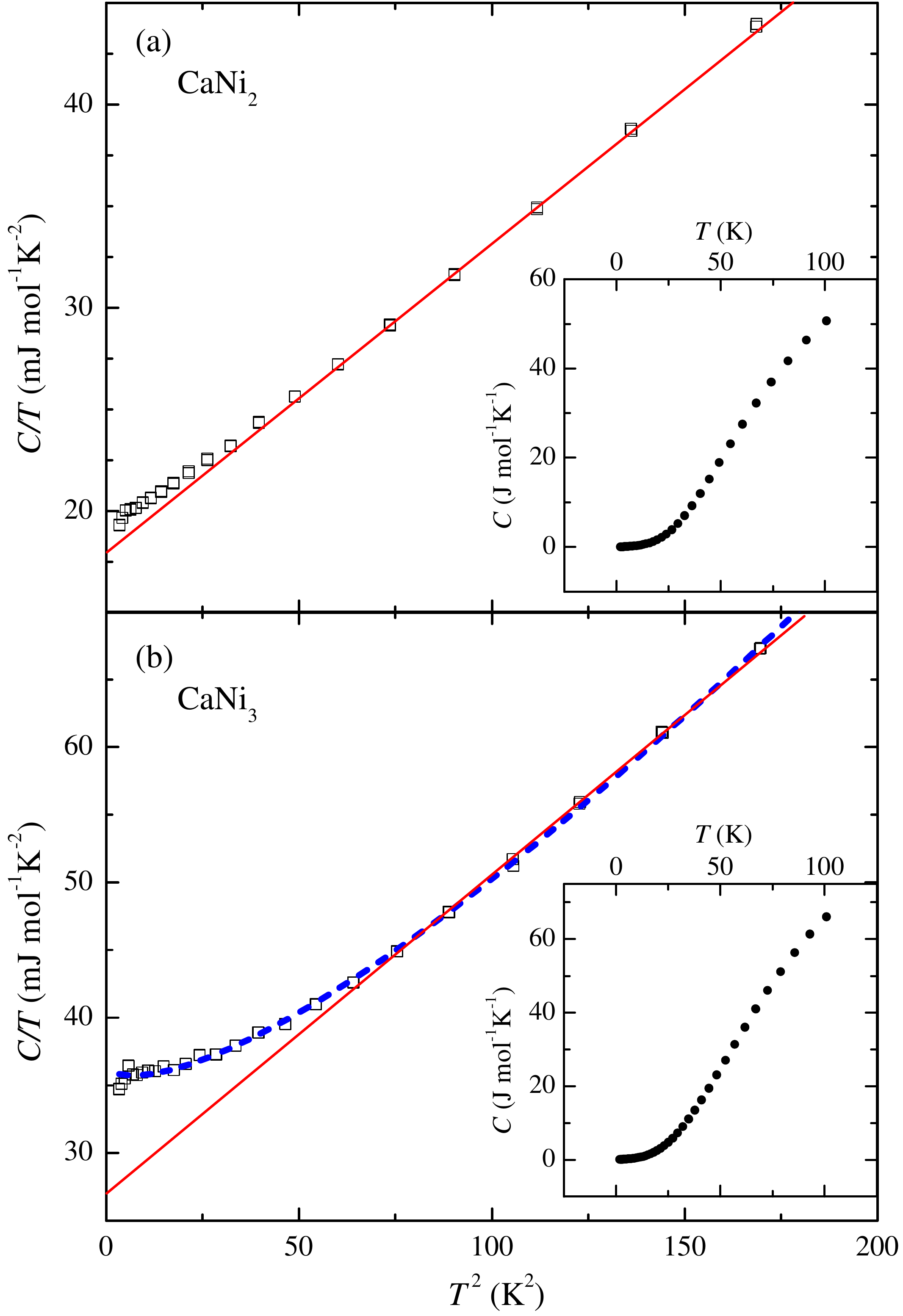}
\caption{(Color online) Specific heat plotted as $C/T$ vs. $T^2$ for a) CaNi$_2$ and b) CaNi$_3$. Additional contributions to the specific heat of an ordinary metal ($C/T = \gamma + \beta T^2$ - indicated by the red lines) are observed below $T \approx 10$\,K. 
Assuming a contribution of spin fluctuations of the form $C_{\rm mag} = A T^3 \mathrm{ln}(T/T_{\rm sf})$ leads to a $T$-dependence shown by the dashed, blue line.
Insets: specific heat as a function of temperature.
}
\label{HC}
\end{figure}

Figure\,\ref{HC} shows the specific heat of CaNi$_2$ (a) and CaNi$_3$ (b) plotted as $C/T$ vs. $T^2$.
Electron and phonon contribution are described by $C = \gamma T + \beta T^3$ and are indicated by red lines in Fig.\,\ref{HC}. A similar Sommerfeld-coefficient of $\gamma = 9.0$\,mJ mol$_{\rm Ni}^{-1}$K$^{-2}$ was found for both compounds (when expressed of in terms per mole nickel).
Additional contributions to the specific heat are observed for $T < 10$\,K. Whereas the deviations are small for CaNi$_2$, there is a significant enhancement of $C/T$ for CaNi$_3$ towards low $T$ most likely associated with FM fluctuations observed in the $\chi(T)$ measurements. 
These magnetic fluctuations contribute to the specific heat by $C_{\rm{mag}} = A T^3 \mathrm{ln}(T/T_{\rm sf})$\,\cite{Tari2003}. The coefficients obtained by fitting the experimental data to the resulting equation $C/T = \gamma_{\rm{mag}} + \beta T^2 + A T^2 \mathrm{ln}(T/T_{\rm sf})$ are $\gamma_{\rm{mag}} = 12.1(2)$\,mJ\,mol$^{-1}_{\rm Ni}$K$^{-2}$, $\beta = 0.080(5)$,mJ\,mol$^{-1}_{\rm Ni}$K$^{-4}$, $A = 0.056(3)$\,mJ\,mol$^{-1}_{\rm Ni}$K$^{-4}$, and $T_{\rm sf} = 18(2)$\,K . The corresponding fit is plotted as dashed blue line in Fig.\,\ref{HC}b. 
Note that the contribution of spin fluctuations to the specific heat is negative for $T < T_{\rm sf}$ corresponding to a shift of entropy to higher $T$ around $T_{\rm sf}$. This is compensated by assuming a higher electron contribution to $C$ at low $T$ ($\gamma_{\rm{mag}} = 12.1$\,mJ\,mol$^{-1}_{\rm Ni}$K$^{-2}$ compared to $\gamma = 9.0$\,mJ\,mol$_{\rm Ni}^{-1}$K$^{-2}$).
Since the deviation of the measured specific heat of CaNi$_2$ from the expected lattice and electron contribution are small, a quantitative estimate of $A$ and $T_{\rm sf}$ is not reliable. 
The specific heat data between $T = 2$ and 100\,K are plotted in the insets of Fig.\,\ref{HC}, showing the expected decrease of slope in the upper $T$-range towards approaching the Dulong-Petit limits (75\,Jmol$^{-1}$K$^{-1}$ and 100\,Jmol$^{-1}$K$^{-1}$, respectively). 

\section{Discussion}
Our results on CaNi$_2$ are in contrast to the findings of Tsvyashchenko \textit{et al.}, claiming weak ferromagnetism for CaNi$_2$ at room-temperature\,\cite{Tsvyashchenko2002}. However, no measured experimental data are presented in Ref.\,\onlinecite{Tsvyashchenko2002}. Therefore, we are not sure whether the discrepancy is due a modification of the structure caused by the high-pressure synthesis used by Tsvyashchenko \textit{et al.} or simply the result of FM nickel impurities (or other second phases) which might be present in their polycrystalline samples.

Next we want to discuss the possibility of local magnetic moments indicated by the Curie-Weiss behavior of the magnetic susceptibility $\chi(T)$. 
In this scenario a large AFM exchange interaction has to be inferred from the observed Weiss-temperatures of $\Theta_W \approx -500$ and $-900$\,K, for CaNi$_2$ and CaNi$_3$, respectively.
Effective moments of $>$1$\,\mu_B$ (as obtained for this compounds) coupled by such strong interactions are expected to order antiferromagnetically above 100\,K. 
However, there is no indication for an AFM ordering in the whole $T$-range investigated.
Therefore, the experimental results are incompatible with a local moment scenario unless the system is assumed to be extremely frustrated (highly unlikely in both of such different structures).
Taking into account that Ni is not known to carry a local moment together with the FM correlations inferred from the low $T$ upturn in $\chi(T)$ (Fig.\,\ref{cani3_mag}a) makes this assumption unlikely.

A more plausible explanation for the increase in $\chi(T)$ towards low $T$ is given by an enhancement of the normally only weakly $T$-dependent terms of the paramagnetic susceptibility of a metal. 
Similar temperature dependence of $\chi(T)$ was observed for elemental Pd\,\cite{Gerhardt1981} and more recently for YFe$_2$Zn$_{20}$ and LuFe$_2$Zn$_{20}$\,\cite{Jia2007,Jia2009} and can be understood in terms of proximity to the Stoner limit.
Thereby, the theoretical description is based on the temperature dependence of the Fermi-Dirac distribution function (within the framework of Stoner-theory) and leads to a qualitative agreement with the experimental data (see e.g. Ref.\,\onlinecite{Zellermann2004}). 

The Stoner factor $Z$, defined by $\chi = \frac{\chi_{\rm para}}{1-Z}$, can be used to quantify the strength of FM correlations (where $\chi$ is the renormalized Pauli susceptibility of a metal, $\chi_{\rm para}$).
$Z$ can be calculated from the experimental data of $\chi(T \rightarrow 0)$ and $\gamma$ by 
$Z = 1 - \frac{3\mu_B}{(\pi k_B)^2}\frac{\gamma}{\chi(T \rightarrow 0)}$ 
(assuming density of states and magnetic susceptibility of the free electron gas).
We find $Z = 0.79$ (CaNi$_2$) and $Z = 0.87$ (CaNi$_3$, $\mu_0H = 1$\,T, $\bm H \perp \bm c$).

For the archetypical Stoner enhanced metal, elemental Pd, the calculation yields $Z = 0.83$
\lbrack $\gamma = 9.45$\,mJ\,mol$^{-1}$K$^{-2}$ (Ref.\,\onlinecite{Stewart1983}) and $\chi(T\,\rightarrow\,0)$ = 9.29 $\cdot10^{-9}$\,m$^3$mol$^{-1}$ (Ref.\,\onlinecite{Gerhardt1981}) of Pd are both similar to the values obtained for CaNi$_2$ and CaNi$_3$\rbrack.
Therefore, at least CaNi$_3$ is even closer to a FM ordered ground state than Pd and approaches the high values of $Z = 0.88$ found in YFe$_2$Zn$_{20}$ and LuFe$_2$Zn$_{20}$\,\cite{Jia2009}.
Taking into account the field dependence of $\chi(T \rightarrow 0)$ of CaNi$_3$ (steeper increase of $M(H)$ for $\mu_0H < 1$\,T, see Fig.\,\ref{cani3_mag}c), even higher Stoner factors are obtained, e.g. $Z = 0.89$ for $\mu_0H = 0.5$\,T.
Using the higher $\gamma_{\rm mag} = 12.1$\,mJ\,mol$^{-1}_{\rm Ni}$K$^{-2}$ for CaNi$_3$, and in doing so assume a larger electron contribution to the specific heat and accordingly a higher density of states at the Fermi-level, still leads to $Z = 0.85$.

A major difference between CaNi$_2$ and CaNi$_3$ and other Stoner-enhanced metals like elemental Pd, YFe$_2$Zn$_{20}$ and LuFe$_2$Zn$_{20}$ is the absence of a maximum in $\chi(T)$ at low $T$.
As shown by Yamada\,\cite{Yamada1993} this maximum is correlated with a possible itinerant electron metamagnetic transition within Ginzburg-Landau theory and occurs when the Landau expansion coefficients ($A,B,C$) of the magnetic part of the free energy, $\Delta F(M) = \frac{1}{2}AM^2 + \frac{1}{4}BM^4 + \frac{1}{6}CM^6$, fulfill the condition $\frac{AC}{B^2} > \frac{5}{28}$. 
This is not the case for CaNi$_2$ and CaNi$_3$ (in the temperature range investigated) and accordingly no indications for a metamagnetic transition were observed in $M(H)$ measurements.

Consistent with FM spin fluctuations inferred from the field-dependence of $\chi(T)$ we observed an upturn in the low temperature specific heat of CaNi$_3$ (Fig.\,\ref{HC}b). 
Similar behavior has been observed in other FM, but close to paramagnetic (Ni$_{0.63}$Rh$_{0.37}$\,\cite{Bucher1967}), or nearly FM itinerant electron systems (TiBe$_2$\,\cite{Stewart1982}). 
It remains an open question why this upturn is absent or significantly less pronounced in the nearly FM itinerant electron systems like Pd, YFe$_2$Zn$_{20}$, and LuFe$_2$Zn$_{20}$. 

Since CaNi$_3$ was found to be very close to a FM ordered ground state and the iso-structural compound YNi$_3$ is a FM system with $T_C = 30$\,K\,\cite{Gignoux1980}, we tried to tune the system by gradually substituting Y for Ca. However, first attempts to synthesize the alloy Ca$_{1-x}$Y$_x$Ni$_3$ failed due to the formation of (Y,Ca)Ni$_5$ and (Y,Ca)$_2$Ni$_7$.
Further efforts to systematically induce a FM transition are underway.
\\

\section{Summary}
Single crystals of CaNi$_2$ and CaNi$_3$ have been grown and characterized by x-ray diffraction and temperature dependent electrical resistivity, magnetization and specific heat measurements.  Both compounds manifest behavior consistent with Stoner enhanced, nearly ferromagnetic Fermi liquids with CaNi$_2$ and CaNi$_3$ having Stoner enhancement factors, Z, of 0.79 and 0.87, respectively.  The low temperature specific heat of CaNi$_3$ shows signatures of ferromagnetic fluctuations consistent with this Stoner enhanced state.

\section{Acknowledgments}
The authors want to thank H. Hodovanets, S. K. Kim, X. Lin, S. M. Sauerbrei, and S. Ran for technical support. S. L. Bud'ko is acknowledged for technical support as well as for fruitful (Kirsche) discussions.  
This work was supported by the U.S. Department of Energy, Office of Basic Energy Science, Division of Materials Sciences and Engineering. The research was performed at the Ames Laboratory. Ames Laboratory is operated for the U.S. Department of Energy by Iowa State University under Contract No. DE-AC02-07CH11358.

\end{document}